\documentclass[twocolumn,showpacs,preprintnumbers,amsmath,amssymb,superscriptaddress]{revtex4}

\usepackage{bm,verbatim}
\usepackage{color,graphicx}

\begin{document}

\preprint{Submitted to {\it Physical Review Letters}}

\title{Super-Alfv\'enic propagation of reconnection signatures and
  Poynting flux during substorms}

\author{M. A. Shay} \email{shay@udel.edu}
\affiliation{Department of Physics \& Astronomy, 217 Sharp Lab,
  University of Delaware, Newark, DE 19716}
\author{J. F. Drake}
\affiliation{University of Maryland, College Park, Maryland 20742}
\author{J. P. Eastwood}
\affiliation{The Blackett Laboratory, Imperial College London, London,
SW7 2AZ, United Kingdom}
\author{T. D. Phan}
\affiliation{Space Sciences Laboratory, University of California,
  Berkeley, California 94720}

\date{\today}

\begin{abstract} The propagation of reconnection signatures and their
  associated energy are examined using kinetic particle-in-cell
  simulations and Cluster satellite observations. It is found that the
  quadrupolar out-of-plane magnetic field near the separatrices
  is associated with a kinetic Alfv\'en wave. For magnetotail
  parameters, the parallel propagation of this wave is
  super-Alfv\'enic $(V_{\parallel} \sim 1500 - 5500\,{\rm km/s})$ and
  generates substantial Poynting flux $(S \sim 10^{-5} - 10^{-4
  }\,{\rm W/m^2})$ consistent with Cluster observations of magnetic
  reconnection. This Poynting flux substantially exceeds that due to
  frozen-in ion bulk outflows and is sufficient to generate white
  light aurora in the Earth's ionosphere. 
  \end{abstract}

\pacs{Valid PACS appear here}

\maketitle

Magnetic reconnection plays an important role in many plasma systems
by releasing large amounts of magnetic energy through the breaking and
reforming of magnetic field lines (e.g., \cite{Yamada10}). During
magnetospheric substorms the global magnetic geometry is
reconfigured\cite{Akasofu64}, releasing magnetotail magnetic energy
and creating intense auroral disturbances. During solar flares,
magnetic energy release in the corona energizes large numbers of
electrons which create hard x-rays when they impact to surface of the
sun (e.g., \cite{Miller97}).

The sudden onset of magnetospheric substorms is believed to be caused
by either a near Earth instability at around $10\,R_e$ downtail~(e.g.,
\cite{Lui96}) or reconnection onset around $20$ to $30\,R_e$~(e.g.,
\cite{Baker96}). Determining the mechanism or mechanisms which are
most relevant requires careful timing studies and has been the subject
of much scrutiny and controversy~(e.g.,
\cite{Angelopoulos08c,Lui09,Angelopoulos09,Kepko09,Nishimura2010}). A
key unanswered question regarding magnetic reconnection, therefore,
regards how fast the released energy and associated signatures
propagate away from the X-line. The propagation of MHD signatures, ion
flows and magnetic disturbances, has been extensively studied in both
substorms (e.g., \cite{Birn99}) and solar flares (e.g.,
\cite{Linton06}), but these mechanisms are limited by the Alfv\'en
speed.  In some substorm events, however, it has been reported that
the time lag between reconnection onset and auroral onset was less
than the Alfv\'en transit time from the reconnection site to the
ionosphere\cite{Angelopoulos08c,Lin09}. It is necessary, therefore, to
determine the nature of the reconnection signal that propagates
fastest away from a reconnection site and its associated
energies. Poynting flux\cite{Wygant00,Keiling03} associated with
kinetic Alfv\'en waves, for example, has been postulated as a possible
energy source for aurora\cite{Lysak04}, with observations of these
waves near magnetotail reconnection sites\cite{Chaston09,Dai09}.

We simulate magnetic reconnection with the kinetic particle-in-cell
code P3D and find that the quadrupolar Hall out-of-plane magnetic
field located near the separatrices is associated with a kinetic
Alfv\'en wave (KAW).  This KAW magnetic field perturbation has a
super-Alfv\'enic parallel propagation speed (using lobe densities),
and is associated with a substantial Poynting flux that points away
from the X-line. This KAW will exist whenever Hall physics is active
in the diffusion region \cite{Rogers01}. Simulation Poynting flux is
consistent with Cluster statistical observations of multiple
magnetotail reconnection events. Scaling to magnetotail and
ionospheric parameters, the transit time of this standing KAW from a
near Earth X-line is on the order of 50 seconds.

{\bf Simulations:} Our simulations are performed with the
particle-in-cell code p3d \cite{Zeiler02}. The results are presented
in normalized units: the magnetic field to the asymptotic value of the
reversed field $B_0$, the density ($n_0$) to the value at the center
of the current sheet minus the uniform background density, velocities
to the Alfv\'en speed $c_A$, lengths to the ion inertial length $d_i$,
times to the inverse ion cyclotron frequency $\Omega_{ci}^{-1}$,
temperatures to $m_i c_A^2$, and Poynting flux to $S_0 =
c_A\,B_0^2/4\pi$. We consider a system periodic in the $x-z$ plane
where flow into and away from the X-line are parallel to
$\mathbf{\hat{z}}$ and $\mathbf{\hat{x}}$, respectively.  The
initial equilibrium consists of two Harris current sheets superimposed
on an ambient population with a uniform density of $0.2$. The
equilibrium magnetic field is given by $B_x=\tanh[(y-L_z/4)/w_0]-
\tanh[(y-3L_z/4)/w_0]-1$, where $w_0$ and $L_z$ are the half-width of
the initial current sheets and the box size.  The electron and ion
temperatures, $T_e = 1/12$ and $T_i = 5/12$, are initially uniform.
The simulations presented here are two-dimensional, {\it i.e.},
$\partial/\partial y = 0$. Reconnection is initiated with a small
initial magnetic perturbation.

We have explored the separatrix structure and reconnection signal with
three different simulations, varying the electron mass as shown in
Table~\ref{table-results}.
\begin{table}
\caption{\label{table-results} Simulation parameters and results:
  $\Delta = $ grid scale, $c = $ light speed, $(L_x,L_z)$ = system size,
$(\lambda,k,V_{sim},S_{sim})$ = properties of
KAW. $(d_{i\ell},d_{e\ell})$ = lobe inertial lengths.}
\begin{tabular*}{8.3cm}{@{\extracolsep{\fill}}cccccccccc}
\hline\hline
$\frac{m_i}{m_e}$ & $\frac{\Delta}{d_i}$ & $\frac{c}{c_A}$ & $\frac{L_x}{d_i}$ & $\frac{L_z}
{d_i}$ & $\frac{\lambda}{d_i}$ 
& $k\,d_{i\ell}$ & $k\,d_{e\ell}$ 
& $\frac{V_{\par sim}}{c_A}$ & $\frac{S_{\par sim}}{S_0}$ \\ \hline
25 & 0.05 & 15 & 204.8 & 102.4 & 7.4 & 1.9 & 0.38 & 2.3 & 0.08\\
100 & 0.025 & 20 & 102.4 & 51.2 & 4.0 & 3.5 & 0.35 & 3.1 & 0.13\\
400 & 0.0125 & 40 & 51.2 & 25.6 & 2.6 & 5.4 & 0.27 & 4.0 & 0.18\\
\hline\hline
\end{tabular*}
\end{table}
These simulations were used for a previous study of
reconnection\cite{Shay07}. As shown in Fig. 1 of
citation\cite{Shay07} the reconnection rate increases
with time, sometimes undergoes a modest overshoot, and approaches a
quasi-steady rate of around $0.15$. 

\begin{figure}
\includegraphics[width=8.6cm]{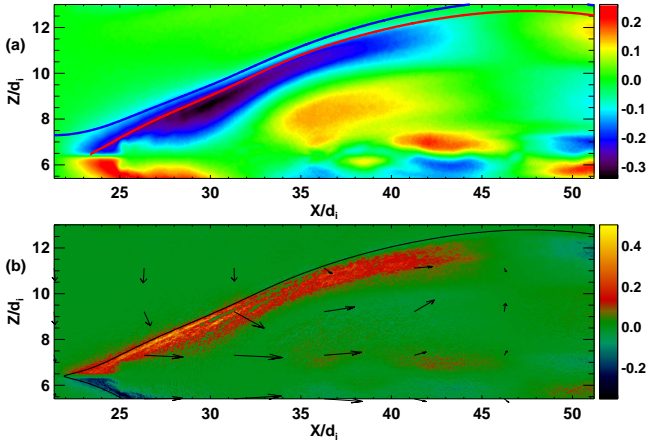}
\caption{\label{2d-overview} 2D overview plots. (a) $B_y$, with red
  and blue magnetic field line segments used in
  Fig.~\ref{wave-propagation}. (b) Parallel poynting flux ${\bf S}
  \cdot \hat{\bf b}_{xz}$ with ion velocity vectors. Black line is
  separatrix magnetic field line.}
\end{figure}

The structure of roughly one quadrant of the reconnection region is
shown in Fig.~\ref{2d-overview}, with the X-line located at
$(x/d_i,z/d_i) = (21.36,6.40)$.
  The large $B_y$ associated with
Hall physics is clearly evident near the separatrix. This magnetic
field is produced by nearly parallel electron flows near the
separatrices, which are strongly super-Alfv\'enic. There is a strong
Poynting flux parallel to the in-plane magnetic field, ${\bf S} \cdot
\hat{\bf b}_{xz}$, where $\hat{\bf b}_{xz} = ({\bf B}_x + {\bf
  B}_z)/\sqrt{B_x^2 + B_z^2}$. It is this Poynting flux which carries
the energy of the first signal of reconnection. Note that there is
little ion flow associated with this $B_y$ and Poynting flux.

In order to gain some handle on the physics governing this $B_y$
structure associated with reconnection, we represent it as a
superposition of linear waves with various $k$ values. The scaling
laws based on this analysis will be shown to be consistent with
simulation properties. Examining Fig.~\ref{2d-overview}a, the quasi-1D
$B_y$ structure is very nearly parallel to the separatrix and thus is
a strongly oblique wave with $k_\parallel \ll k_\perp$. As a starting
point, we use the two-fluid analysis from previous
studies\cite{Rogers01,Drake08}, and analyze the branch of waves
associated with Alfv\'en waves and kinetic Alfv\'en waves. For the
simulation parameters used in this study and noting also that $k d_e
\lesssim 1,$ with $d_e$ the electron skin depth: \begin{equation}
\label{dispersion} \frac{\omega^2}{k_\parallel^2} = \frac{c_{A}^2}{D}
\left[ 1 + \left(\frac{k^2 d_i^2}{D} \right) \frac{c_s^2}{c_A^2/D +
c_s^2} \right], \end{equation} with $D = 1+k^2 d_e^2$ and $c_s^2 =
(T_e + T_i)/m_i.$ Note that for highly oblique waves, the parallel
group velocity is equal to the parallel phase velocity.

\begin{figure}
\includegraphics[width=8.6cm]{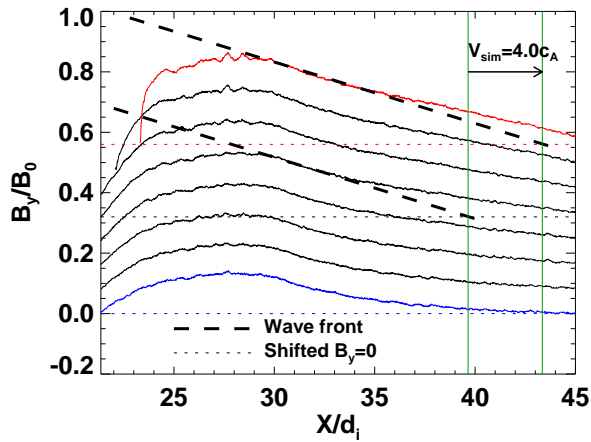}
\caption{\label{wave-propagation} $B_y$ plotted along magnetic field
  lines separated by $\Delta \psi = 0.05\,B_0d_i,$ which represents
  $\Delta t = 0.31\,\Omega_{ci}^{-1}$. Vertical offset of each plot is
  $0.08\,B_0$. Wave fronts for $\Delta \psi = 0.2\,B_0d_i$ and
  $0.35\,B_0d_i$ shown as dashed lines, with respective $B_y = 0$
  shown as horizontal dotted lines. Blue and red $B_y$ plots taken 
  along field line segments shown in
  Fig.~\ref{2d-overview}a. Wave front intersections with $B_y = 0$
  denoted with vertical green lines.}  \end{figure}

{\bf Simulation KAW:} This analysis uses quasi-steady reconnection to
study the properties of the separatrix kinetic Alfv\'en wave
(KAW). This is necessary because the KAW is so fast that the
quadrupolar field associated with it very quickly fills the whole
simulation domain making velocity measurements due to direct time
variation impossible; the high speed of the KAW is most likely why its
propagation velocities have been largely ignored by previous studies
of collisionless reconnection, although other properties of the
quadrupolar field have been extensively examined through simulations,
satellite observations, and laboratory experiments (\cite{Yamada10},
and references therein). During steady reconnection, magnetic field
lines convect along the inflow ($z$) direction, reconnect, and then
flow outwards. The propagation velocity of the KAW can be measured by
changing frames to one moving with that inflowing magnetic field
line. Since the reconnection is steady, the time difference between
two magnetic field lines is the difference in flux between the two
lines over the reconnection rate, i.e., $\Delta t = \Delta \psi / E_r
= $ $ \Delta \psi / (\,\partial/\partial t (\psi_{\rm xline} - \psi_{\rm
  oline})\,),$ where $\psi$ is defined such that ${\bf B} = \hat{\bf
  y} \times \nabla \psi + B_y\,\hat{\bf y}$. By examining the KAW
$B_y$ at different $\psi$ values, therefore, one can determine the
propagation speed of the KAW. An example of this analysis for the
$m_e/m_i=1/400$ case is shown in Fig.~\ref{wave-propagation}, which
shows the variation of $B_y$ along magnetic field lines (lines of
constant $\psi$).
 For clarity,
two representative magnetic field line segments colored red and blue
are shown in Fig.~\ref{2d-overview}a, and the $B_y$ plots taken along
them are colored the same. Each $B_y$ plot represents a $\Delta \psi =
0.05\,B_0d_i$, and each successive plot has been offset $0.08\,B_0$
along the vertical from the previous one. The evolution of the $B_y$
is not characterized by a simple propagation. First, the peak value of
$B_y$ increases with time. Second, the dispersive nature of KAWs also
leads to multiple velocities associated with the $B_y$ structure. The
location where $B_y = 0$ propagates at the peak KAW speed, $V_{\rm
  peak} \approx C_{se} = $ $\sqrt{(T_e + T_i)/m_e} \approx
14\,c_A$. However, there is little Poynting flux associated with this
velocity. Instead, we focus on the propagation of the main $B_y$
signal by finding the velocity of the wave front. The two dashed lines
in Fig.~\ref{wave-propagation} have the same slope and denote the wave
front in two of the curves separated by $\Delta \psi = 0.15\,B_0d_i.$
The propagation velocity of the x-intercept of this slope (shown as
vertical green lines) is calculated to be $4.0\,c_A,$ which is
substantially less than the peak parallel KAW speed.  The measured
values are shown as $V_{sim}/c_A$ in Table~\ref{table-results}.

\begin{figure}
\includegraphics[width=8.6cm]{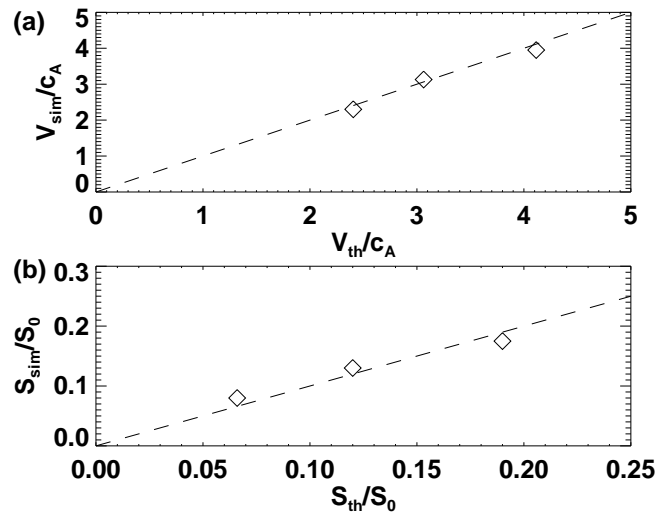}
\caption{\label{compare-theory-sim} Simulation values versus
  theoretical predictions for (a) KAW propagation speed and (b) Poynting
  flux.}
\end{figure}

It is critical to determine if this propagation velocity is consistent
with the kinetic Alfven wave predictions of
Eq.~(\ref{dispersion}). First, the $k$ values associated with this
$B_y$ Hall field must be determined. Vertical slices of the Poynting
flux ${\bf S} \cdot \hat{\bf b}_{xz}$ were analyzed at the locations
of the wave front ($m_i/m_e = [25,100,400]$, $x/d_i = [170.0, 90.0,
35.0]$). The magnitude of this Poynting flux is shown in
Table~\ref{table-results} as $S_{sim}$. The width at half-max of the
Poynting flux was measured and used to determine the primary $k =
2\pi/\lambda$ value for the KAW. As an example, for the $m_i/m_e=400$
case, the half max was $\delta = 0.65\,d_i,$ yielding $\lambda \approx
2.6\,d_i$. The standing KAW wave is located close to the separatrices,
so the simulation lobe plasma values are used to determine parameters
($B \approx 1.0, n \approx 0.2, T_i + T_e \approx 0.5$), which yields
$k d_{i\ell} \approx 5.4,$ where the ``$\ell$'' denotes lobe
values. The resulting $\lambda, k d_{i\ell},$ and $k d_{e\ell}$ are
shown in Table~\ref{table-results}. Plotting the velocities predicted
from Eq.~(\ref{dispersion}) versus the simulation measured velocities
yields excellent agreement, as shown in
Fig.~\ref{compare-theory-sim}a.


Associated with this Hall structure are electron beams and significant
Poynting flux. The super-Alfv\'enic electron beams are associated with
the parallel currents which create the quadrupolar $B_y$. A
theoretical prediction for the Poynting flux can be determined for
comparison with simulation values. We use ${\bf S} \cdot \hat{\bf
  b}_{xz} \approx S_x = (c/4\pi)({\bf E} \times {\bf B})_x \approx
-(c/4\pi)E_z B_y$. The normal Hall electric field is due to the
frozen-in electron flow, which dominates over the ion flow, giving
$E_z \approx V_{ex} B_y/c \approx -J_x B_y/(nec) \approx -B_y
B_y'/(4\pi ne)$ with $B_y' = \partial B_y/\partial z$. Substituting
gives $S_x \approx B_y B_y' \,c_{Ay}\, d_i/4\pi$, where $c_{Ay} =
B_y/\sqrt{4\pi m_i n}$. Note that the integrated KAW Poynting flux is
independent of the width of the KAW. As with the KAW velocity
determination, $n$ is the lobe density with $B_y \approx 0.25$
consistent with simulation values. Comparison of the theoretical
Poynting fluxes with simulation values also yields excellent
agreement, as seen in Fig.~\ref{compare-theory-sim}b. Note that this
KAW Poynting flux substantially exceeds the Poynting flux associated
with the ion bulk flow away from the X-line: $S_{x,\,\rm ion} \approx
c_A B_z^2/4\pi,$ since $B_z^2 \approx 0.01\,B_0^2 \ll B_y^2$.

\begin{figure}
\includegraphics[width=8.6cm]{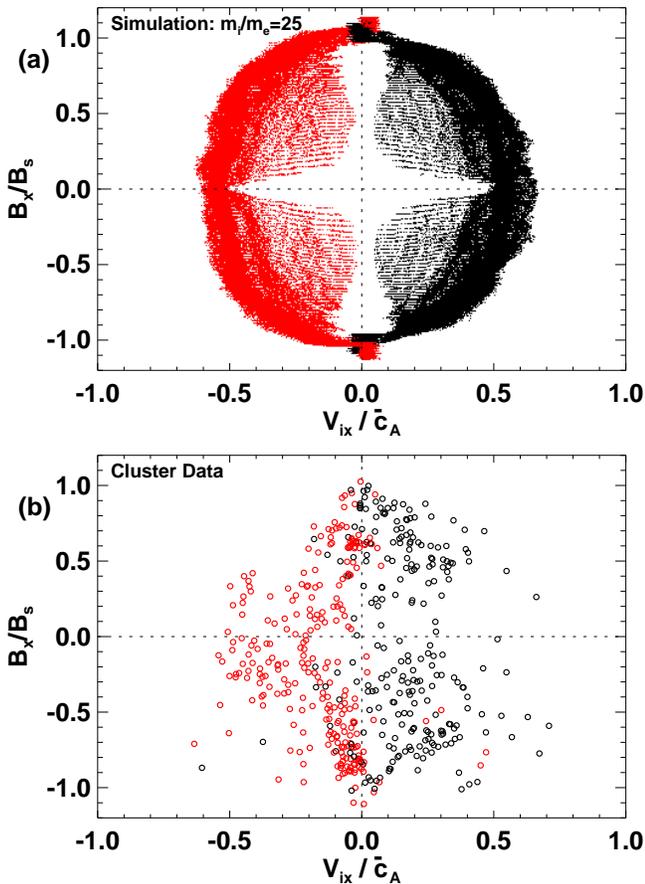}
\caption{\label{compare-sim-satellite} (a) Simulation and (b) Cluster
  Observations: Poynting flux $S_x/\bar{S} > 0.02$ scatter plot in
  $(B_x/B_s,V_{ix}/\bar{c}_A)$ plane (red: Tailward $S_x/\bar{S}$,
  black: Earthward $S_x/\bar{S}$). Simulation values in (a) are
  smaller circles due to large number of data points. Normalizations
  described in text.}  \end{figure}

{\bf Comparisons with Satellite Data:} A statistical study of
reconnection events has been performed previously\cite{Eastwood2010b},
where magnetotail reconnection crossings with correlated Geocentric
Solar Magnetospheric (GSM) $B_z$ and $V_{ix}$ reversals were
selected. In that study\cite{Eastwood2010b}, comparisons with
simulations were made by renormalizing data using magnetic fields just
upstream of the separatrices ($B_s$) and densities in the ion outflow
region ($n_{\rm out}$), yielding normalization velocity $\bar{c}_A =
B_s/\sqrt{4\pi m_i n_{\rm out}}$ and Poynting flux $\bar{S} =
\bar{c}_A\, B_s^2/4\pi.$ Using these normalizations, the Poynting flux
from this Cluster data set is compared with data from the $m_i/m_e=25$
case. For the simulation data, the normalization values used were $B_s
= 0.8$ and $n_{\rm out} = 0.2$. The simulation sub-region used was a
rectangle roughly centered on the X-line with length approximately
$35\,d_i$ and height approximately $13\,d_i$, using
$n=0.2$. Fig.~\ref{compare-sim-satellite} shows this comparison, where
only normalized $S_x/\bar{S} > 0.02$ is plotted. Tailward
$S_x/{\bar{S}}$ is shown in red and Earthward $S_x/\bar{S}$ is shown
in black.  
The bounds of the simulation and Cluster data are similar, being limited
to $\vert V_{ix}\vert \lessapprox 0.7$ and $\vert B_x\vert \lessapprox
1.0$. The separatrix KAW structure is present in both plots in the
region of large $B_x$ and nearly zero $V_{ix}$. Both datasets show a
strong correlation in the sign of $V_{ix}$ and $S_x,$ implying that
the Poynting flux points away from the X-line. However, for small
$V_{ix}$ and larger $B_x$ there is some anti-correlation which
corresponds to ion flow towards the X-line just outside the
separatrices.  Both data sets show significant $S_x$ for small $\vert
B_x\vert$ and larger negative $V_{ix}$, which is associated with the
very long outflow jet of super-Alfv\'enic electrons seen in
simulations with kinetic electrons\cite{Shay07,Karimabadi07} and
satellite observations\cite{Phan07}. There is an asymmetry, however,
in the satellite data along $V_{ix}$ not present in the simulations,
with only negative (tailward) $S_x/\bar{S}$ having significant values
for $B_x \approx 0$ and finite $V_{ix}$. Some possible explanations
are: (1) In most of the events, the satellite was initially tailward
of the X-line and then crossed to the Earthward side, so Earthward
flows represent more developed X-lines. (2) The obstacle presented by
the strong Earth's dipole field could create back pressure and lead
to outflow asymmetries at the X-line. Or (3) 3D effects lead to this asymmetry.

{\bf Predictions for the Magnetotail: } The KAW associated with the
quadrupolar $B_y$ propagates at a super-Alfv\'enic speed and carries
significant Poynting flux. To assess its importance for the
magnetosphere, we use the following typical
parameters\cite{Angelopoulos08c}: $B \approx 20\,{\rm nT},\, n \approx
0.1\,{\rm cm}^{-3},\, T_e \approx 300\,{\rm eV}$, and $T_i \approx
1\,{\rm keV}$. As the KAW propagates large distances in the
magnetotail, it is quite probable that the $k$ associated with it will
decrease owing to the dispersive nature of KAWs. Taking the simulation
$k d_e \approx 0.3$ to be the maximum expected $k$, we take $k d_i
\approx 1$ to be the minimum $k$ because at this $k$ the KAWs are no
longer dispersive. As is found in the simulations, we use $B_y/B_{\rm
  lobe} \approx 0.25$. These values yield the following ranges of
parameters associated with the KAW: $V_\parallel \sim 1500-5500\,{\rm
  km/s},$ $S \sim 0.7\cdot 10^{-5}-9\cdot 10^{-5}\,{\rm W/m^2}$. For an X-line located
$20\,R_e$ downtail from the Earth, the predicted propagation time is
$\Delta t \sim 25 - 85\,{\rm sec}$, which is substantially less than
the Alfv\'en transit time $(\sim 250\,{\rm sec})$ for the same distance.

An important question remains as to whether this KAW energy will be
able to propagate to the Earth's ionosphere and create aurora. In the
simulations (largest $L_x \approx 10\,R_e$ and $\Delta t \approx
50\,s$ using simulation lobe parameters), the KAW propagates all the
way to the edge of the simulation, but the limited length scale as
well of lack of a dipole geometry make exact estimation of the wave
attenuation impossible. This is an important question currently under
study.  Assuming parallel propagation of the Poynting flux so that it
stays on the same magnetic flux tube, the Poynting flux in the
ionosphere $S_{\rm ion}$ would be: $S_{\rm ion} \sim (B_{\rm
ion}/B_{\rm lobe}) S_{\rm lobe} \sim 10^3\, S_{\rm lobe}$. Reducing
this flux by a factor of ten as an estimate of attenuation yields:
$S_{\rm ion} \sim 10^2\, S_{\rm lobe} \sim 0.7\cdot 10^{-3} - 9\cdot
10^{-3}\,{\rm W/m^2}$, which is still on the order of or greater than
the $10^{-3}\,{\rm W/m^2} = 1\,{\rm ergs/cm^2s}$ necessary to create a
white light aurora.

{\bf Acknowledgments} This work was supported by NASA grant
NNX08AM37G, NSF grant ATM-0645271, and the STFC grant ST/G00725X/1 at
Imperial College London. Computations were carried out at the National
Energy Research Scientific Computing Center. The authors thank
V. Angelopoulos, A. T. Y. Lui, T. Nishimura, L. Lyons, L. Kepko, and
R. Lysak for helpful discussions.


\begin{thebibliography}{25}
\expandafter\ifx\csname natexlab\endcsname\relax\def\natexlab#1{#1}\fi
\expandafter\ifx\csname bibnamefont\endcsname\relax
  \def\bibnamefont#1{#1}\fi
\expandafter\ifx\csname bibfnamefont\endcsname\relax
  \def\bibfnamefont#1{#1}\fi
\expandafter\ifx\csname citenamefont\endcsname\relax
  \def\citenamefont#1{#1}\fi
\expandafter\ifx\csname url\endcsname\relax
  \def\url#1{\texttt{#1}}\fi
\expandafter\ifx\csname urlprefix\endcsname\relax\def\urlprefix{URL }\fi
\providecommand{\bibinfo}[2]{#2}
\providecommand{\eprint}[2][]{\url{#2}}

\bibitem[{\citenamefont{{Yamada} et~al.}(2010)\citenamefont{{Yamada},
  {Kulsrud}, and {Ji}}}]{Yamada10}
\bibinfo{author}{\bibfnamefont{M.}~\bibnamefont{{Yamada}}} et al.,
 \bibinfo{journal}{Rev. Modern Phys.} \textbf{\bibinfo{volume}{82}},
  \bibinfo{pages}{603} (\bibinfo{year}{2010}).

\bibitem[{\citenamefont{Akasofu}(1964)}]{Akasofu64}
\bibinfo{author}{\bibfnamefont{S.~I.} \bibnamefont{Akasofu}},
  \bibinfo{journal}{Planet. Space Sci.} \textbf{\bibinfo{volume}{12}},
  \bibinfo{pages}{273} (\bibinfo{year}{1964}).

\bibitem[{\citenamefont{Miller et~al.}(1997)\citenamefont{Miller,
  Cargill, Emslie, Holman, Dennis, LaRosa, Winglee, Benka, and
  Tsuneta}}]{Miller97} \bibinfo{author}{\bibfnamefont{J.~A.}
  \bibnamefont{Miller}} et al., \bibinfo{journal}{J. Geophys. Res.}
  \textbf{\bibinfo{volume}{102}}, \bibinfo{pages}{14631}
  (\bibinfo{year}{1997}).

\bibitem[{\citenamefont{Lui}(1996)}]{Lui96}
\bibinfo{author}{\bibfnamefont{A.~T.~Y.} \bibnamefont{Lui}},
  \bibinfo{journal}{J. Geophys. Res.} \textbf{\bibinfo{volume}{101}},
  \bibinfo{pages}{13067} (\bibinfo{year}{1996}).

\bibitem[{\citenamefont{{Baker} et~al.}(1996)\citenamefont{{Baker},
  {Pulkkinen}, {Angelopoulos}, {Baumjohann}, and {McPherron}}}]{Baker96}
\bibinfo{author}{\bibfnamefont{D.~N.} \bibnamefont{{Baker}}} et al.,
 \bibinfo{journal}{J. Geophys. Res.}
  \textbf{\bibinfo{volume}{101}}, \bibinfo{pages}{12975}
  (\bibinfo{year}{1996}).

\bibitem[{\citenamefont{{Angelopoulos}
  et~al.}(2008)\citenamefont{{Angelopoulos}, {McFadden}, {Larson}, {Carlson},
  {Mende}, {Frey}, {Phan}, {Sibeck}, {Glassmeier}, {Auster}
  et~al.}}]{Angelopoulos08c}
\bibinfo{author}{\bibfnamefont{V.}~\bibnamefont{{Angelopoulos}}} et
al.,
 \bibinfo{journal}{Science}
  \textbf{\bibinfo{volume}{321}}, \bibinfo{pages}{931} (\bibinfo{year}{2008}).

\bibitem[{\citenamefont{Lui}(2009)}]{Lui09}
\bibinfo{author}{\bibfnamefont{A.~T.~Y.} \bibnamefont{Lui}},
  \bibinfo{journal}{Science} \textbf{\bibinfo{volume}{324}},
  \bibinfo{pages}{1391} (\bibinfo{year}{2009}).

\bibitem[{\citenamefont{{Angelopoulos}
  et~al.}(2009)\citenamefont{{Angelopoulos}, {McFadden}, {Larson}, {Carlson},
  {Mende}, {Frey}, {Phan}, {Sibeck}, {Glassmeier}, {Auster}
  et~al.}}]{Angelopoulos09}
\bibinfo{author}{\bibfnamefont{V.}~\bibnamefont{{Angelopoulos}}} et al.,
 \bibinfo{journal}{Science}
  \textbf{\bibinfo{volume}{324}}, \bibinfo{pages}{1391} (\bibinfo{year}{2009}).

\bibitem[{\citenamefont{{Kepko} et~al.}(2009)\citenamefont{{Kepko},
  {Spanswick}, {Angelopoulos}, {Donovan}, {McFadden}, {Glassmeier}, {Raeder},
  and {Singer}}}]{Kepko09}
\bibinfo{author}{\bibfnamefont{L.}~\bibnamefont{{Kepko}}} et al.,
 \bibinfo{journal}{Geophys. Res. Lett.} \textbf{\bibinfo{volume}{36}},
  \bibinfo{pages}{24104} (\bibinfo{year}{2009}).

\bibitem[{\citenamefont{{Nishimura} et~al.}(2010)\citenamefont{{Nishimura},
  {Lyons}, {Zou}, {Angelopoulos}, and {Mende}}}]{Nishimura2010}
\bibinfo{author}{\bibfnamefont{Y.}~\bibnamefont{{Nishimura}}} et al.,
 \bibinfo{journal}{J. Geophys. Res.} \textbf{\bibinfo{volume}{115}},
  \bibinfo{pages}{7222} (\bibinfo{year}{2010}).

\bibitem[{\citenamefont{{Birn} et~al.}(1999)\citenamefont{{Birn}, {Hesse},
  {Haerendel}, {Baumjohann}, and {Shiokawa}}}]{Birn99}
\bibinfo{author}{\bibfnamefont{J.}~\bibnamefont{{Birn}}} et al.,
 \bibinfo{journal}{J. Geophys. Res.} \textbf{\bibinfo{volume}{104}},
  \bibinfo{pages}{19895} (\bibinfo{year}{1999}).

\bibitem[{\citenamefont{{Linton} and {Longcope}}(2006)}]{Linton06}
\bibinfo{author}{\bibfnamefont{M.~G.} \bibnamefont{{Linton}}} \bibnamefont{and}
  \bibinfo{author}{\bibfnamefont{D.~W.} \bibnamefont{{Longcope}}},
  \bibinfo{journal}{ApJ} \textbf{\bibinfo{volume}{642}}, \bibinfo{pages}{1177}
  (\bibinfo{year}{2006}).

\bibitem[{\citenamefont{{Lin} et~al.}(2009)\citenamefont{{Lin}, {Frey},
  {Mende}, {Mozer}, {Lysak}, {Song}, and {Angelopoulos}}}]{Lin09}
\bibinfo{author}{\bibfnamefont{N.}~\bibnamefont{{Lin}}} et al.,
 \bibinfo{journal}{J. Geophys. Res.} \textbf{\bibinfo{volume}{114}},
  \bibinfo{pages}{12204} (\bibinfo{year}{2009}).

\bibitem[{\citenamefont{{Wygant} et~al.}(2000)\citenamefont{{Wygant},
  {Keiling}, {Cattell}, {Johnson}, {Lysak}, {Temerin}, {Mozer}, {Kletzing},
  {Scudder}, {Peterson} et~al.}}]{Wygant00}
\bibinfo{author}{\bibfnamefont{J.~R.} \bibnamefont{{Wygant}}} et al.,
 \bibinfo{journal}{J. Geophys. Res.}
  \textbf{\bibinfo{volume}{105}}, \bibinfo{pages}{18675}
  (\bibinfo{year}{2000}).

\bibitem[{\citenamefont{{Keiling} et~al.}(2003)\citenamefont{{Keiling},
  {Wygant}, {Cattell}, {Mozer}, and {Russell}}}]{Keiling03}
\bibinfo{author}{\bibfnamefont{A.}~\bibnamefont{{Keiling}}} et al.,
 \bibinfo{journal}{Science}
  \textbf{\bibinfo{volume}{299}}, \bibinfo{pages}{383} (\bibinfo{year}{2003}).

\bibitem[{\citenamefont{Lysak and Song}(2004)}]{Lysak04}
\bibinfo{author}{\bibfnamefont{R.~L.} \bibnamefont{Lysak}} \bibnamefont{and}
  \bibinfo{author}{\bibfnamefont{Y.}~\bibnamefont{Song}}, in
  \emph{\bibinfo{booktitle}{Substorms 7: {P}roceedings of the 7th International
  Conference on Substorms}}, edited by
  \bibinfo{editor}{\bibfnamefont{T.}~\bibnamefont{Pulkinnen}} \bibnamefont{and}
  \bibinfo{editor}{\bibfnamefont{N.}~\bibnamefont{Ganushkina}}
  (\bibinfo{publisher}{Finnish Meteorological Institute},
  \bibinfo{year}{2004}), p.~\bibinfo{pages}{81}.

\bibitem[{\citenamefont{{Chaston} et~al.}(2009)\citenamefont{{Chaston},
  {Johnson}, {Wilber}, {Acuna}, {Goldstein}, and {Reme}}}]{Chaston09}
\bibinfo{author}{\bibfnamefont{C.~C.} \bibnamefont{{Chaston}}} et al.,
 \bibinfo{journal}{Phys. Rev. Lett.} \textbf{\bibinfo{volume}{102}},
  \bibinfo{pages}{015001} (\bibinfo{year}{2009}).

\bibitem[{\citenamefont{Dai}(2009)}]{Dai09}
\bibinfo{author}{\bibfnamefont{L.}~\bibnamefont{Dai}}, Ph.D. thesis,
  \bibinfo{school}{University of Minnesota} (\bibinfo{year}{2009}).

\bibitem[{\citenamefont{Rogers et~al.}(2001)\citenamefont{Rogers, Denton,
  Drake, and Shay}}]{Rogers01}
\bibinfo{author}{\bibfnamefont{B.~N.} \bibnamefont{Rogers}} et al.,
 \bibinfo{journal}{Phys. Rev. Lett.} \textbf{\bibinfo{volume}{87}},
  \bibinfo{pages}{195004} (\bibinfo{year}{2001}).

\bibitem[{\citenamefont{Zeiler et~al.}(2002)\citenamefont{Zeiler, Biskamp,
  Drake, Rogers, Shay, and Scholer}}]{Zeiler02}
\bibinfo{author}{\bibfnamefont{A.}~\bibnamefont{Zeiler}} et al.,
 \bibinfo{journal}{J. Geophys. Res.} \textbf{\bibinfo{volume}{107}},
  \bibinfo{pages}{1230} (\bibinfo{year}{2002}).

\bibitem[{\citenamefont{Shay et~al.}(2007)\citenamefont{Shay, Drake, and
  Swisdak}}]{Shay07}
\bibinfo{author}{\bibfnamefont{M.~A.} \bibnamefont{Shay}} et al.,
 \bibinfo{journal}{Phys. Rev. Lett.} \textbf{\bibinfo{volume}{99}},
  \bibinfo{pages}{155002} (\bibinfo{year}{2007}).

\bibitem[{\citenamefont{Drake et~al.}(2008)\citenamefont{Drake, Shay, and
  Swisdak}}]{Drake08}
\bibinfo{author}{\bibfnamefont{J.~F.} \bibnamefont{Drake}} et al.,
 \bibinfo{journal}{Phys. Plasmas} \textbf{\bibinfo{volume}{15}},
  \bibinfo{pages}{042306} (\bibinfo{year}{2008}).


\bibitem[{\citenamefont{Eastwood et~al.}(2010)\citenamefont{Eastwood, Phan,
  {\O}ieroset, and Shay}}]{Eastwood2010b}
\bibinfo{author}{\bibfnamefont{J.~P.} \bibnamefont{Eastwood}} et al.,
 \bibinfo{journal}{J. Geophys. Res.} \textbf{\bibinfo{volume}{115}},
  \bibinfo{pages}{A08215} (\bibinfo{year}{2010}).

\bibitem[{\citenamefont{Karimabadi et~al.}(2007)\citenamefont{Karimabadi,
  Daughton, and Scudder}}]{Karimabadi07}
\bibinfo{author}{\bibfnamefont{H.}~\bibnamefont{Karimabadi}} et al.,
 \bibinfo{journal}{Geophys. Res. Lett.} \textbf{\bibinfo{volume}{34}},
  \bibinfo{pages}{L13104}, (\bibinfo{year}{2007}).

\bibitem[{\citenamefont{Phan et~al.}(2007)\citenamefont{Phan, Drake, Shay,
  Mozer, and Eastwood}}]{Phan07}
\bibinfo{author}{\bibfnamefont{T.~D.} \bibnamefont{Phan}} et al.,
 \bibinfo{journal}{Phys. Rev. Lett.} \textbf{\bibinfo{volume}{99}},
  \bibinfo{pages}{255002} (\bibinfo{year}{2007}).

\end{thebibliography}

\end{document}